\documentclass[copyright]{eptcs}

\providecommand{\event}{ICE 2010} 

\providecommand{\firstpage}{1}

\def\titlerunning{A theory of desynchronisable closed loop systems}
\def\authorrunning{H.Beohar \& P.J.L.Cuijpers}

\usepackage{amsmath,amscd,amsbsy}
\usepackage{amssymb}
\usepackage{amsthm}
\usepackage{procalg}
\usepackage{stackrel}
\usepackage{subfigure}
\usepackage[numbers]{natbib}
\usepackage{graphicx}
\usepackage{graphics}

\usepackage{tikz}
\usetikzlibrary{shapes,arrows,automata}
\usetikzlibrary{shapes.symbols}
\usetikzlibrary{shapes.misc}
\usetikzlibrary{calc}

\theoremstyle{plain}
\newtheorem{theorem}{Theorem}[section]
\newtheorem{corollary}[theorem]{Corollary}
\newtheorem{lemma}[theorem]{Lemma}

\theoremstyle{definition}
\newtheorem{definition}[theorem]{Definition}

\newtheorem{proposition}[theorem]{Proposition}

\newcommand{\act}{\ensuremath{A}}
\newcommand{\actlabel}{\ensuremath{\mathit{Act}}}

\newcommand{\ebag}{\ensuremath{\varepsilon}}
\newcommand{\block}{\ensuremath{H}}
\newcommand{\hide}{\ensuremath{I}}
\newcommand{\prc}{\ensuremath{\mathbb{P}}}
\renewcommand{\comm}{\ensuremath{?\mkern-7mu!}}
\newcommand*\com{\mathop{\raisebox{-1pt}{\normalfont\Large\comm}}}

\newcommand{\longtwoheadleftarrow}{\longleftarrow\hspace{-1.4em}\leftarrow\hspace{.2em}}
\newcommand{\lla}[2]{\ensuremath{\stackrel[]{\mkern+5mu#1}{\longtwoheadleftarrow_{#2}}}}

\newcommand{\bag}{\ensuremath{B}}
\newcommand{\nat}{\ensuremath{\mathbb{N}}}
\newcommand{\well}{\ensuremath{W}}
\newcommand{\hidepi}{{\hide_{P}^{?}}}
\newcommand{\hidepo}{{\hide_{P}^{!}}}
\newcommand{\reach}{\ensuremath{\mathit{Reach}}}

\newcommand{\vecm}{\ensuremath{\vec{\mu}}}
\newcommand{\vecn}{\ensuremath{\vec{\nu}}}

\renewcommand{\nabla}[1]{\abstr{\hat{\hide}}{\encap{\block\cup\hat{\block}}{#1}}}

\newcounter{xcount}
\newcommand{\inct}{T\arabic{xcount}\addtocounter{xcount}{1}}

\title{A theory of desynchronisable closed loop systems}
\author{Harsh Beohar
\institute{Formal methods group\\
Department of Mathematics and Computer Science\\
Eindhoven university of technology, The Netherlands}
\email{H.Beohar@tue.nl}
\and
Pieter Cuijpers
\institute{Formal methods group\\
Department of Mathematics and Computer Science\\
Eindhoven university of technology, The Netherlands}
\email{P.J.L.Cuijpers@tue.nl}
}

\begin{document}
\maketitle

\begin{abstract}
The task of implementing a supervisory controller is non-trivial, even though different theories exist that allow automatic synthesis of these controllers in the form of automata. One of the reasons for this discord is due to the asynchronous interaction between a plant and its controller in implementations, whereas the existing supervisory control theories assume synchronous interaction. As a consequence the implementation suffer from the so-called inexact synchronisation problem. In this paper we address the issue of inexact synchronisation in a process algebraic setting, by solving a more general problem of refinement. We construct an asynchronous closed loop system by introducing a communication medium in a given synchronous closed loop system. Our goal is to find sufficient conditions under which a synchronous closed loop system is branching bisimilar to its corresponding asynchronous closed loop system.
\end{abstract}

\section{Introduction}\label{sec:introduction}

The task of implementing a supervisory controller is non-trivial, even though different theories exist that allow automatic synthesis of these controllers in the form of automata. One of the reasons for this discord is due to the asynchronous interaction between a plant and its controller in implementations, whereas the existing supervisory control theories assume synchronous interaction. We elaborate on this mismatch by first introducing some terminology that is often used in supervisory control theory \citep{RW:1987}.

Supervisory control theory provides an automatic synthesis of a supervisor that controls a plant in such a way that a corresponding requirement (legal behaviour) is achieved. In supervisory control theory terminology,
\begin{itemize}
\item the model that is to be controlled is known as \textit{plant},
\item the model that specifies the requirement is known as \textit{specification},
\item the model that forces the plant to meet the specification by interacting with it is known as \textit{supervisor} or \textit{controller}.
\item the interaction between a plant and its supervisor is known as \textit{closed-loop behavior}.
\end{itemize}
The closed loop behaviour in supervisory control theory is realized by synchronous parallel composition. Informally, it allows a plant and a supervisor to synchronise on common events while other events can happen independently.

One of the main drawbacks while implementing the interaction between a plant and its supervisor, synthesised by supervisory control theory, is inexact synchronization \citep{fabian}. In practical industrial applications, the interaction between a plant and its supervisor is not synchronous but rather asynchronous. Due to the synchronous parallel composition used in supervisory control theory, the interaction between a plant and its supervisor is strict. By strict, we mean that either plant or supervisor has to wait for the other party while synchronising. To overcome this problem it is important to study asynchronous communication between a plant and its supervisor where communications are delayed in buffers.

\citeauthor{balemiphdt} was the first to consider the inexact synchronisation problem, and the solutions given in his PhD thesis \citep{balemiphdt} were in the domain of automata theory. In \citep{balemiphdt}, an \textit{input-output} interpretation was given between a plant and its supervisor and a special delay operator was introduced to model the delay in communication between the plant and the supervisor. Moreover, for this setup the existence of a supervisor in the presence of delays was also shown in \citep{balemiphdt}. It was required that the output actions from a plant can occur asynchronously, while the output actions from a supervisor must occur synchronously \citep{async-imp}. In \citep{async-imp} this requirement was relaxed. Furthermore, necessary and sufficient conditions were also provided for the existence of a controller under bounded delay between a plant and its supervisor.

The solutions provided in \citep{balemiphdt,async-imp} construct a new supervisor under the presence of bounded delay, which is a computationally expensive procedure. To circumvent this, we present sufficient conditions on a synchronous closed loop system under which the asynchronous closed loop system constructed from it, is a refinement of the given synchronous closed loop system. Moreover, the technique developed in this paper is independent of the size of buffers used. However, we do not analyse the computational complexities associated with the sufficient conditions presented in this paper.

In this paper, we reformulate the inexact synchronisation problem as a problem of refinement in the process algebra TCP \citep{acpbook}. The synchronous closed loop system can be considered as a specification with the asynchronous closed loop system as its implementation. If the given synchronous closed loop system and its corresponding asynchronous closed loop system are branching bisimilar \citep{Glabeek90}, then the asynchronous closed loop system is said to be a refinement of its corresponding synchronous closed loop system. Note that we do not compute an additional supervisor under the presence of delays, instead we assume a given plant and its supervisor. Thus, we solve a refinement problem instead of solving a control synthesis problem.

In the past, the idea of solving a refinement problem was studied \citep{Fischer96,HHJ90,diccs}, but different setups (in comparison with the current paper) were used in these studies. These studies were motivated by the so-called ``Foam-rubber wrapper'' principle \citep{udding}, borrowed from the field of delay insensitive circuits. Mathematically, it states that ``a process and the same process connected with buffers are equivalent''. In \citep{Fischer96}, the foam-rubber wrapper principle was also studied in the context of the parallel composition and it was shown that an extra condition is required to preserve this principle. In brief, we have a different architecture for the asynchronous closed loop system in this paper and we study the components in the asynchronous closed loop system conjointly, in order to capture desynchronisability.


\subsection{Architecture}\label{subsec-arch}

This paper is a result of the pre-study carried out in \citep{prDCL}, where four construction methods are proposed to construct an asynchronous closed loop system from its corresponding synchronous one. In this subsection, we introduce the architecture of an asynchronous closed loop system, discuss the reasonability of using a bag as a buffer and describe one of the abstraction schemes that will be used throughout this paper. We elucidate on these points in the upcoming paragraphs.

\begin{figure}
\centering
  \includegraphics[width=5cm]{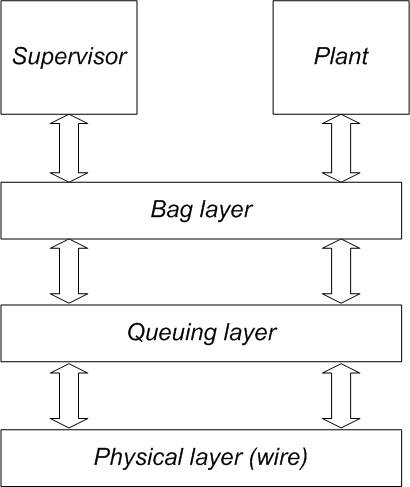}\\
  \caption{Asynchronous closed loop system in practice.}\label{fig:layer}
\end{figure}

An asynchronous closed loop system can be constructed by introducing a buffer between a plant and its supervisor in order to decouple the synchronisation of events between the two. In practice, the buffering mechanism is realised by the interactions of different layers (also known as protocol stack) as shown in Figure~\ref{fig:layer}. In theory, various authors \citep{Fischer96,HHJ90,MousaviDate04} have abstracted from the interaction of different layers by using data structures based on a particular level of abstraction. For example, to model delay insensitive (DI) circuits, which are at a lower level of abstraction (physical layer), wires are used as a buffering mechanism \citep{diccs}. On the other hand to model data flow networks, which are at a higher level of abstraction (in comparison to DI circuits), queues are used as a buffering mechanism \citep{HHJ90}. In this paper, we are interested in studying the asynchronous interaction in a closed loop system at an even higher level of abstraction by having a unique queue for every message. Thus, a queue stores only one type of unique message and all queues are allowed to run concurrently without interacting with one another. Such interleaving queues are equivalent to a bag modulo strong bisimulation. Hence, we use a bag as the buffering mechanism in this paper.

\begin{figure}\centering
  \subfigure[Construction method M1.]{\label{cm-m1}
\includegraphics[width=0.35\textwidth]{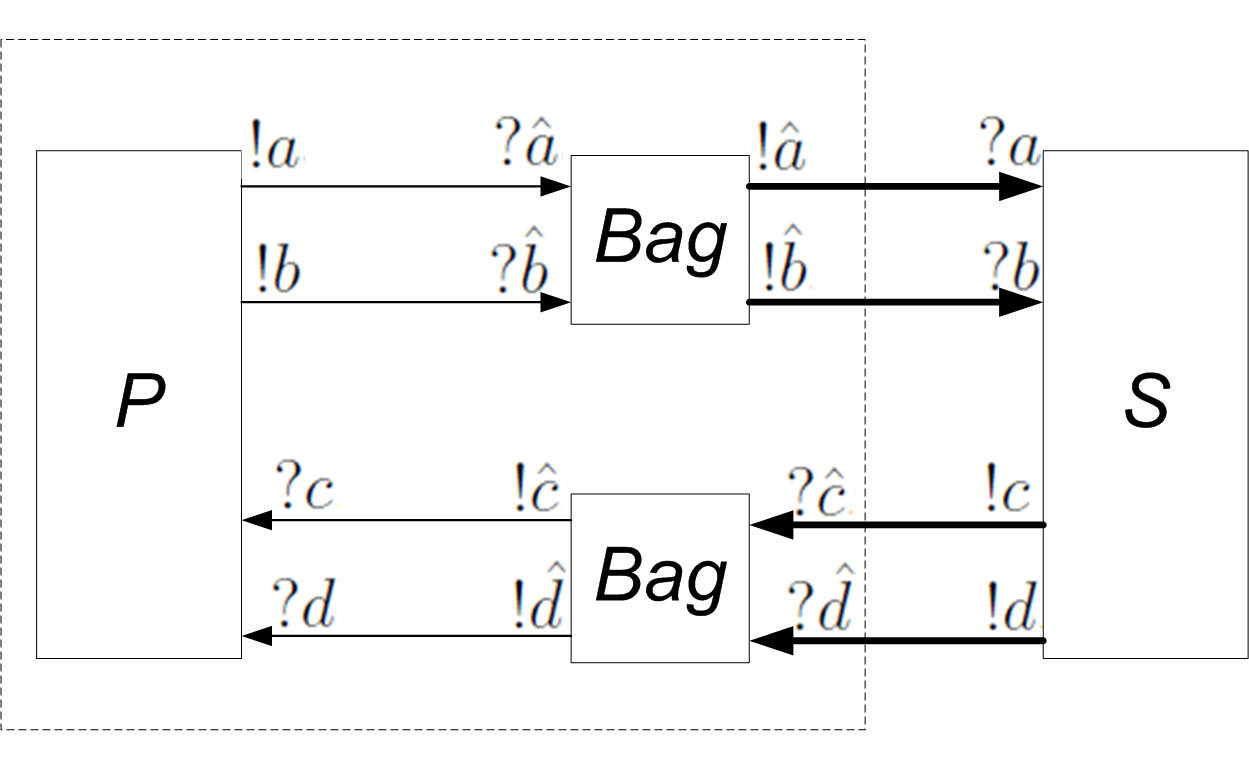}}
\hspace{1cm}  \subfigure[Construction method M2.]{\label{cm-m2}
\includegraphics[width=0.35\textwidth]{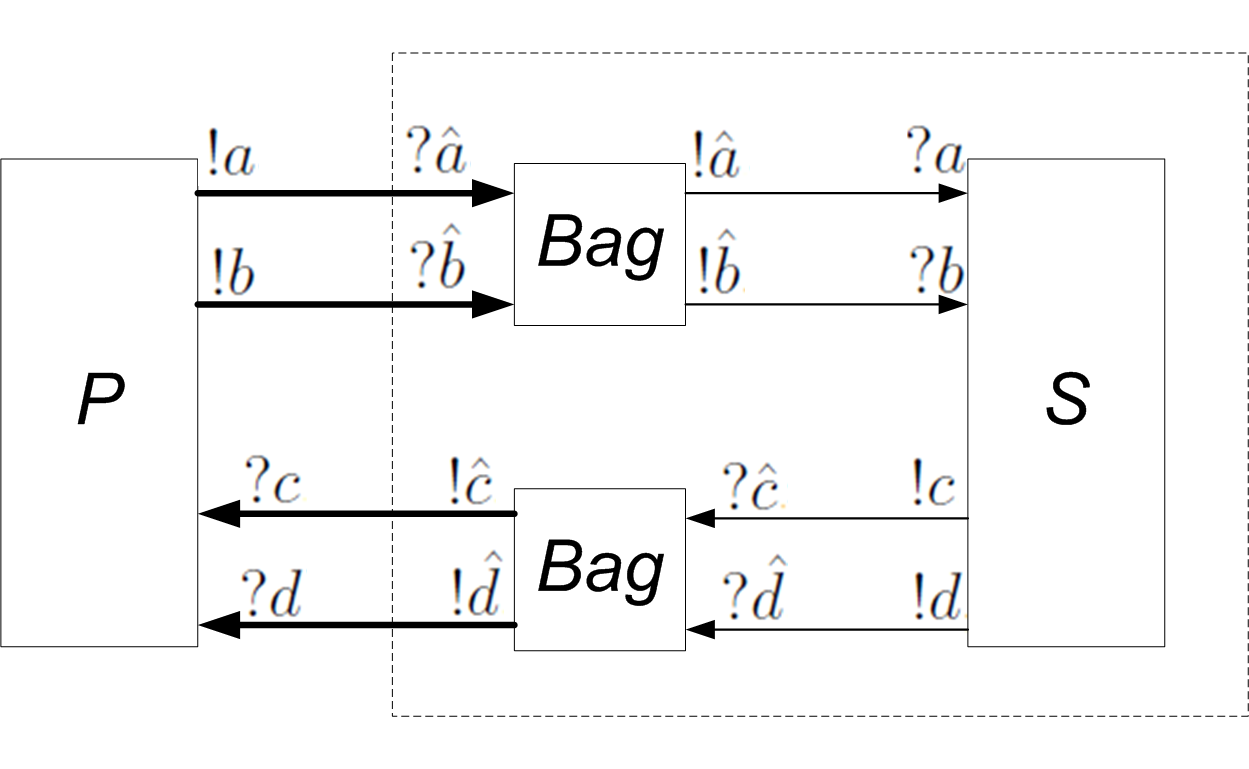}
  }
\hspace{1cm} \subfigure[Construction method M3.]{\label{cm-m3}
\includegraphics[width=0.35\textwidth]{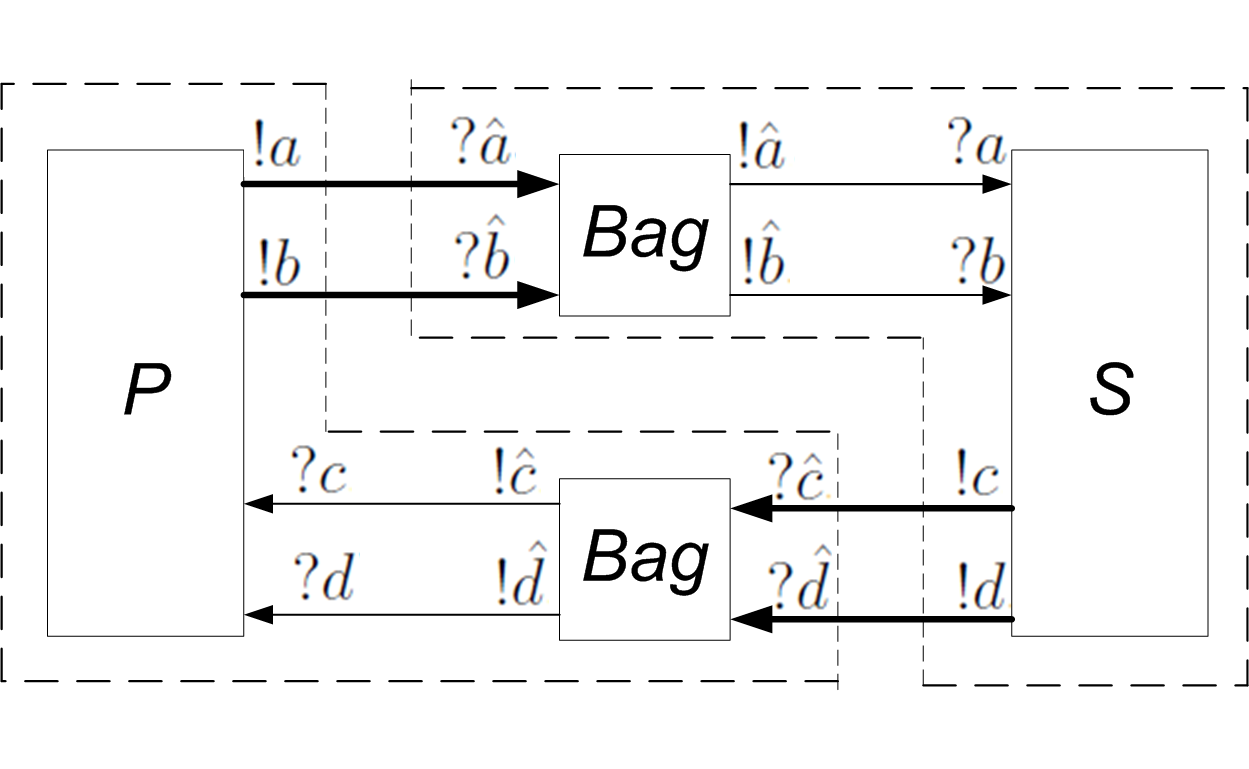}
  }
\hspace{1cm} \subfigure[Construction method M4.]{\label{cm-m4}
\includegraphics[width=0.35\textwidth]{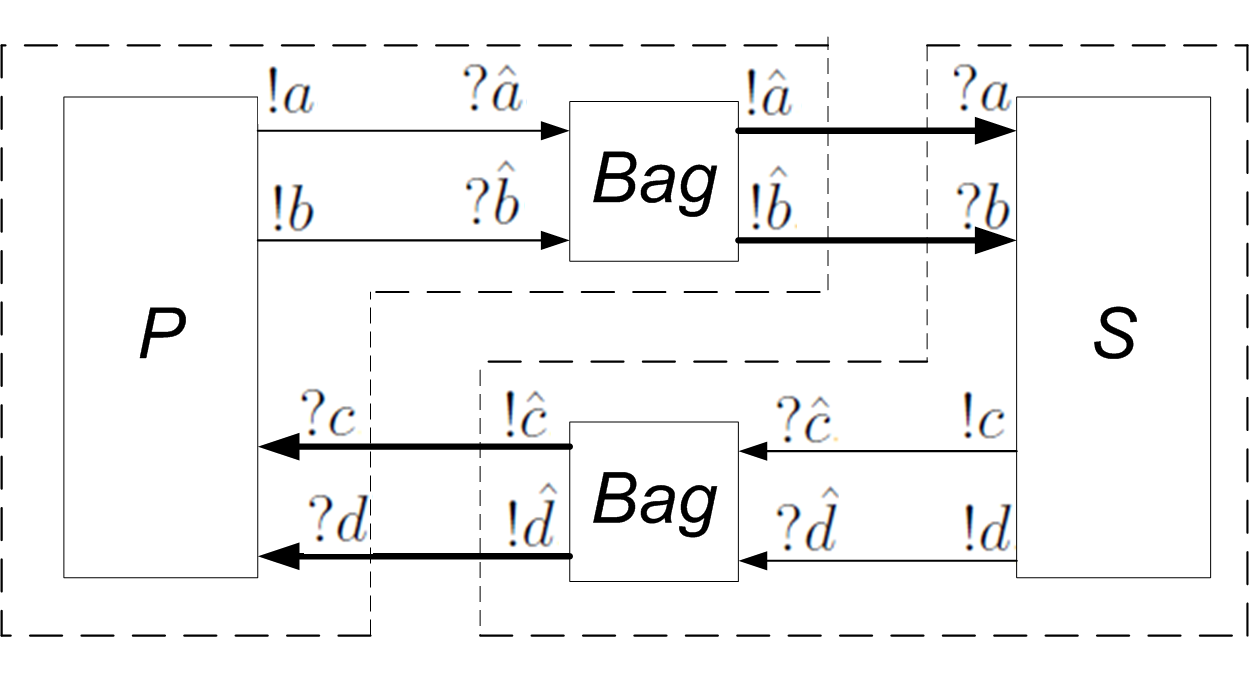}
  }
\caption{Different ways to construct an asynchronous closed loop system.}\label{cm}
\end{figure}

It is obvious that upon introduction of the bag as a buffer, the asynchronous closed loop system contains interactions that are not present in the synchronous closed loop system. However, to relate these two closed loop systems by a branching bisimulation relation \citep{Glabeek90}, it is necessary to hide some interactions or define a suitable abstraction scheme. In principle, a synchronous closed loop system can be converted into an asynchronous closed loop system by introducing bags with the following abstraction schemes:
\begin{itemize}
\item[M1.] by introducing bags between a plant and its supervisor such that \textit{the interaction between plant and bag is hidden} (see Figure~\ref{cm-m1}).
\item[M2.] by introducing bags between a plant and its supervisor such that \textit{the interaction between supervisor and bag is hidden} (see Figure~\ref{cm-m2}).
\item[M3.] by introducing bags between a plant and its supervisor such that \textit{the communication among the input actions of both plant and supervisor with bags are hidden} (see Figure~\ref{cm-m3}).
\item[M4.] by introducing bags between a plant and its supervisor such that \textit{the communication among the output actions of both plant and supervisor with bags are hidden} (see Figure~\ref{cm-m4}).
\end{itemize}
In Figure~\ref{cm}, thick lines are used to show the visible interaction and thin lines are used to show the invisible interaction. The notation $!a$ means `send action $a$' and $?a$ means `receive action $a$'. In this paper, we develop the theory for the construction method M1 (see Section~\ref{sec:synctoasync} for the rationale behind this choice) and leave other construction methods as open for future study. Moreover, the techniques presented in this paper are restricted to reactive systems (so, no termination).

\subsection{Outline}
The remainder of this paper is organized as follows. In Section~\ref{sec:background}, we start our exposition by defining the overall background required for this paper. Section~\ref{sec:introtorw} provides a brief introduction to supervisory control theory with respect to our setup. In Section~\ref{sec:synctoasync}, the construction method M1 is defined formally with its abstraction scheme. In Section~\ref{sec:DCL}, we give the formal definition of a desynchronisable closed loop system with the conditions that are sufficient for desynchronisability. Finally, in Section~\ref{sec:discussion} we present the conclusions and propose some directions for future research.

\section{Background}\label{sec:background}

In this section, we define the basic notations and definitions that will be used throughout this article. Let $\actlabel$ be a set of action names. We use symbols $a,b,c,\dots$ to range over the set $\actlabel$. Then we define the following actions for an action label $a\in\actlabel$,
\begin{itemize}
\item $!a$: send action label $a$.
\item $?a$: receive action label $a$.
\item $\com a$: communicated action label $a$.
\end{itemize}
Let $\act$ denote the set of all possible actions that are defined as, $\act=\{!a,?a,\com a\}_{a\in\actlabel}.$ The variables $x,y,z,\dots$ are used to denote elements from set $\act$ when the information about the type of action is irrelevant. The set of all process terms (denoted by $\prc$) is then defined by the following grammar. The constant $\dl$ is a process term that cannot perform any action, i.e. it can only \emph{deadlock}. A unary operator $x.\_$ for each action $x\in\act$ is introduced in the TCP syntax, denoting an \emph{action prefix}. Intuitively, the process term $x.p$ performs the action $x$ and then behaves as the process $p$. The binary operator $\altc$ denotes the \emph{alternative composition or choice} between any two processes. The encapsulation operator $\encap{\block}{\_}$, \emph{blocks execution of actions} from $\block$ while enduring the execution of other actions from $\act\setminus\block$. The abstraction operator $\abstr{\hide}{\_}$ \emph{renames} the actions from $\hide$ to $\tau$, and leaves other actions unchanged.

\begin{center}
$\begin{array}{lcll}
    \prc& ::= & \dl & \mbox{deadlock process}\\
    &\mid & x.\prc & \mbox{action prefix}\\
    &\mid & \prc\altc \prc & \mbox{alternative composition}\\
    &\mid & \prc \merge_\gamma \prc & \mbox{parallel composition}\\
    &\mid & \encap{H}{\prc} & \mbox{action encapsulation, where }H\subseteq\act\\
    &\mid & \abstr{I}{\prc} &\mbox{abstraction (hiding of actions), where }I\subseteq\act\\
    &\mid & \mathcal{R} & \mbox{recursive definition}
\end{array}$
\end{center}

In the remainder of this paper, we assume that the symbols $P,R,S,p,p',s,s'\dots$ range over the set $\prc$. We fix the capital letters $P,R,S$ for processes associated with supervisory control theory. Note that we also use the alphabet operator $\alpha$ and renaming operator $\rho$ from TCP algebra for technical reasons, but not for modeling purposes. The empty process $\emp$ is not defined because we are interested in modeling only reactive systems. The notation $\mathcal{R}$ denotes a recursion definition by a set of pairs $\{X_0=t_0,\dots, X_m=t_m\}$ where $X_i$ denotes a recursion variable and $t_i$ the process term defining it. The parallel composition operator is parameterized with a communication function $\gamma:\act\times\act\rightarrow\act$ such that $\gamma(?a,!a)=\gamma(!a,?a)=\com a\enspace.$

The semantic domain of the process terms is a transition system (See \citep{acpbook} for details), which is achieved by the so-called SOS rules \citep{SOS}. For the sake of completeness, we give the SOS rules of the operators used here in the Appendix~\ref{osrules}.

\begin{definition}
A \emph{transition system} over a set of actions $\act$ is a set $Q$ of states, equipped with a transition relation $\rightarrow\subseteq Q\times\act\cup\{\tau\}\times Q$. The action $\tau\not\in\act$ denotes the invisible action. In the semantics of TCP, $Q$ is usually taken to be the set of process terms, i.e., $Q=\prc$, and the initial state of a process is defined as the process term itself. \qed
\end{definition}

As mentioned in the introduction, we use branching bisimulation to relate a synchronous closed loop system and its corresponding asynchronous closed loop system in which $\tau$ actions are present. The presence of $\tau$ actions in an asynchronous closed loop system will become evident in Section~\ref{sec:synctoasync}. We write the transitive closure of the transition relation $\rightarrow$ as $\twoheadrightarrow$. The symbol $\equiv$ is used to denote syntactical equivalence between process terms. The shorthand notation $q\steps{\tau^{*}}q'$ is defined as $q\equiv q_0\step{\tau}\dots \step{\tau}q_n\equiv q'$ for all $q_i\in Q$ with $i\in[0,n],n\geq0$.

\begin{definition}\label{def-bb}
A binary relation $\Phi\subseteq Q\times Q$ is called a \emph{branching bisimulation relation} \citep{acpbook,Glabeek90} iff:
\begin{itemize}
\item $\forall q,q_1,q',x.\Big[(q,q')\in\Phi\wedge q\step{x}q_1\Rightarrow\exists q_1',q_2'.\big[q'\steps{\tau^{*}}q_1'\step{x}q_2'\wedge (q,q_1')\in\Phi \wedge (q_1,q_2')\in\Phi\big]\Big]$.
\item $\forall q,q_1,q'.\Big[(q,q')\in\Phi\wedge q\step{\tau}q_1\Rightarrow(q_1,q')\in\Phi\vee
    \exists q_1',q_2'.\big[q'\steps{\tau^{*}}q_1' \step{\tau}q_2'\wedge (q,q_1')\in\Phi\wedge (q_1,q_2')\in\Phi\big]\Big]$.
\item $\forall q,q',q_1',x.\Big[(q,q')\in\Phi\wedge q'\step{x}q_1'\Rightarrow\exists q_1,q_2.\big[q\steps{\tau^{*}}q_1\step{x}q_2\wedge (q_1,q')\in\Phi \wedge (q_2,q_1')\in\Phi\big]\Big]$.
\item $\forall q,q',q_1'.\Big[(q,q')\in\Phi\wedge q'\step{\tau}q_1'\Rightarrow(q,q_1')\in\Phi\vee
    \exists q_1,q_2.\big[q\steps{\tau^{*}}q_1 \step{\tau}q_2\wedge (q_1,q')\in\Phi\wedge (q_2,q_1')\in\Phi\big]\Big]$.
\end{itemize}
Let $q,q'\in Q$ be the initial states of processes $p,p'\in\prc$, respectively. Two processes $p$ and $p'$ are said to be branching bisimilar (denoted as $p\bbisim p'$) iff there exists a branching bisimulation relation $\Phi$ such that there initial states $q,q'$ are related, i.e. $(q,q')\in\Phi$. \qed
\end{definition}

Note that in the absence of $\tau$ actions, branching bisimulation coincides with strong bisimulation. The phenomena of the occurrences of redundant silent steps can be formulated by the following notion of $\tau$-inertness \citep{confluence}.

\begin{definition}
Let $p\in\prc$ be an arbitrary process. A process $p$ is said to be $\tau$\emph{-inert} with respect to $\bbisim$ iff for all states $q$ of the transition system (generated by operational rules) of $p$ it holds that $q\step{\tau}q'\Rightarrow q\bbisim q'$ where, $q'\in Q$. \qed
\end{definition}

The essence of the above definition is that an inert $\tau$ action does not affect the future choices of a process modulo branching bisimulation. In Section~\ref{sec:DCL}, we also show that an asynchronous closed loop system constructed from a synchronous closed loop system satisfying Definitions~\ref{def-wellposed}, \ref{def-diamond} and \ref{def-reordering} is always $\tau$-inert with respect to $\bbisim$.

\section{Supervisory control theory}\label{sec:introtorw}

In this section, we give a brief introduction to supervisory control theory and define its fundamental entities in our setup. The basic entity (a plant, or a supervisor, or a requirement) in the supervisory control theory is deterministic. Furthermore, the proof of main Theorem~\ref{thm:char} requires the fact that a given synchronous closed loop system is also deterministic. Therefore, we now introduce the term deterministic process.

\begin{definition}\label{def:determinism}
A process $p\in\prc$ is called a \emph{deterministic process} iff
for all states $q$ of the transition system (generated by the operational rules) of $p$ and for all $x\in\act$ it holds that $q\step{x}q_1\wedge q\step{x}q_2\Rightarrow q_1\equiv q_2$ where, $q_1,q_2\in Q$.\qed
\end{definition}

In supervisory control theory, plants and supervisors are allowed to perform events that are divided into two disjoint subsets: \emph{controllable} and \emph{uncontrollable} events. The idea behind this partition is that the supervisor can enable or disable controllable events so that the closed loop behavior is equivalent to the requirement. The supervisor can observe but cannot influence uncontrollable events. In this paper, we follow the input-output interpretation \cite{balemiphdt} between a plant and its supervisor; wherein the uncontrollable events are outputs from a plant to a supervisor and the controllable events are outputs from a supervisor to a plant. Thus, processes that model plants or supervisors must have distinct (because of the above partition) input and output actions in its alphabet. Such processes are called input-output processes.

\begin{definition}
The set of input actions for an arbitrary process $p\in\prc$ is denoted by $\alpha^?(p)$ and is defined as $\alpha^?(p)\triangleq\{?a\mid ?a\in\alpha(p)\}$. Similarly, the set of output actions (denoted by $\alpha^!(p)$) is defined as $\alpha^!(p)\triangleq\{!a\mid !a\in\alpha(p)\}$. A process $p$ is called an \emph{input-output process} iff \[\alpha^?(p)\cap\alpha^!(p)=\emptyset\wedge \encap{\hide}{p}\bbisim p \wedge \tau\not\in\alpha(p)\]
where, $\hide=\{\com a\mid a\in\actlabel\}.$\qed
\end{definition}

The condition $\encap{\hide}{p}\bbisim p$ ensures that an input-output process does not contain communicated actions in its alphabet. This is because bags are introduced to buffer both input and output events of an input-output process $p\in\prc$. So if communicated actions are allowed in the specification of the process $p$ then, the information whether the action $\com a$ is an input or an output action of the process $p$ is unknown.

We now define the three basic entities in the supervisory control theory in our setup. A plant $P\in\prc$ is a deterministic and an input-output process. Similarly, a supervisor is a deterministic and an input-output process. A requirement is a process specifying the legal interaction that should occur while the plant and its supervisor are interacting such that a required task (for which the supervisor is synthesised) is completed. Thus, a requirement is a deterministic process $R\in\prc$ such that, $$\encap{\block}{R}\bbisim R\wedge \tau\not\in\alpha(R),$$
where $\block=\{!a,?a \mid a\in\actlabel\}$. This condition ensures that a requirement process only contains communicated actions in its alphabet.

Now, we can state the control problem as follows: given a plant $P$ and a requirement $R$, find a supervisor $S$ such that, \[\encap{\block}{P\merge_\gamma S}\bbisim R.\]
In this paper, we are not interested in how this supervisor is computed and rather assume that we are provided with a solution to the above equation. The goal of this paper is then to find certain conditions on the given synchronous closed loop system such that it is desynchronisable. Note that in supervisory control theory the control problem is based on language equivalence, but branching bisimilarity coincides with language equivalence in the presence of determinism and in the absence of $\tau$ actions. However, we use branching bisimulation because the asynchronous closed loop systems as constructed in the next section are always nondeterministic. In brief, this cause of nondeterminism is due to the abstraction of interactions between a plant and the buffer.

\section{From synchrony to asynchrony}\label{sec:synctoasync}

In the previous section, we formally defined a plant $P$, a supervisor $S$ and a requirement $R$ in our setup. Now, we extend our setup in accordance with the architecture of Subsection~\ref{subsec-arch}, to model asynchronous communication by introducing two bags between a given plant and its supervisor; one bag that contains input actions of $P$ and another one that contains output actions of $P$. Next we define a multiset and some operations over multisets that are necessary for the definition of a bag.

A multiset $\xi$ over the set of communicated actions $\hide$ is a tuple $(I,\kappa)$ where $\kappa:I\rightarrow\nat$ is the corresponding multiplicity function. We write the empty multiset as $\ebag$, which is defined as $(\emptyset,\kappa_0)$, where $\kappa_0:\emptyset\rightarrow 0$ is the \emph{zero function}.

\begin{definition}
Let $\xi=(I,\kappa)$ be a multiset over the set $\hide$.
\begin{itemize}
\item The predicate $\in'$ is used to denote an element that \emph{belongs} to a multiset. It is defined as $\com a\in'\xi\triangleq \com a\in I\wedge \kappa(\com a)>0$.
\item The operator $\oplus$ is used to denote an \emph{addition} of an element to a multiset. It is defined as $\xi\oplus\com a\triangleq(I',\kappa')$ where,\newline $I'=\left\{
                                         \begin{array}{ll}
                                           I, & \hbox{if $\com a\in I$} \\
                                           I\cup\{\com a\}, & \hbox{if $\com a\not\in I$}
                                         \end{array}
                                       \right.
$ and
$\kappa'(x)=\left\{
                      \begin{array}{ll}
                       \kappa(\com a)+1, & \hbox{if $x=\com a\wedge\com a\in I$} \\
                       1, & \hbox{if $x=\com a\wedge\com a\not\in I$}\\
                       \kappa(x) & \hbox{if $x\neq\com a\wedge x\in I$}
                        \end{array}
                         \right.
    $.
\item The operator $\ominus$ is used to denote a \emph{removal} of an element from a multiset. It is defined as $\xi\ominus\com a\triangleq(I',\kappa')$ where,\newline
    $I'=\left\{
         \begin{array}{ll}
           I, & \hbox{$\kappa(\com a)>1$} \\
           I\setminus \{\com a\}, & \hbox{$\kappa(\com a)=1$}
         \end{array}
       \right.
    $ and
    $\kappa'(x)=\left\{
               \begin{array}{ll}
                 \kappa(\com a)-1, & \hbox{if $x=\com a\wedge\kappa(\com a)>0$} \\
                 \kappa(x), & \hbox{if $x\neq\com a\wedge x\in I$}
               \end{array}
             \right.
$.\qed
\end{itemize}
\end{definition}

For each $x\in\act$ we define a new element $\hat{x}$. Let $\hat{\act}$ denote the set of new elements of the form $\hat{x}$.
Similarly we assume that there exists auxiliary hidden and blocking sets $\hat{\hide}$ and $\hat{\block}$, respectively.

\begin{definition}{(\textbf{Bag}).}\label{bagdef}
Let $n>0$ be a natural number representing the size of a bag process. Let $\ebag$ denote the empty multiset and $\xi$ denote a multiset of communicated actions (i.e. the actions that are decorated with the symbol $\com$). Then a \emph{bag process} over a set of actions $A_1\subseteq\act$ of size $n$ is defined in the following way.
\begin{eqnarray*}
B_{A_1}^{n}(\ebag,0)&=&\sum_{?a\in A_1}?\hat{a}.B_{A_1}^{n}(\ebag\oplus\com a,1)\enspace,\\
B_{A_1}^{n}(\xi,i) &=& \sum_{\com a\in'\xi}!\hat{a}.B_{A_1}^{n}(\xi\ominus\com a,i-1) + \sum_{?b\in {A_1}}?\hat{b}.B_{A_1}^{n}(\xi\oplus\com b,i+1)\quad \mbox{for every $0<i<n$}\enspace,\\
B_{A_1}^{n}(\xi,n) &=& \sum_{\com a\in'\xi}!\hat{a}.B_{A_1}^{n}(\xi\ominus\com a,n-1)\enspace.
\end{eqnarray*}\qed
\end{definition}
The above definition is bounded with variable $n$ that not only helps in modeling a realistic asynchronous implementation (as they contain buffers with finite memory). In contrast, it also aids in modeling an asynchronous implementation having buffers of infinite size, i.e., when $n=\infty$. Notation, we denote the two interleaving bags as, $$\bag^{m,n}[\ebag,\ebag]\triangleq\bag_{A_1}^{m}(\ebag)\merge
\bag_{A_2}^{n}(\ebag)$$
where, $A_1=\alpha^?(P)$, $A_2=\alpha^!(P)$ and $m>0$ ($n>0$) denotes the size of bag associated with input (output) actions of the plant $P$. Furthermore, the sets $A_1$ and $A_2$ denote the set of input and output actions of the plant $P$, respectively.

Next we formally define an abstraction scheme that implements the construction method M1. Informally, it decorates the interaction between a plant and the two interleaving bags with the symbol $\hat{\_}$, indicating such interactions are to be made hidden. We write the asynchronous closed loop system as $\abstr{\hat{\hide}}{\encap{\block\cup\hat{\block}}{P\merge_{\gamma'} \bag^{m,n}[\ebag,\ebag] \merge_{\gamma'}S}}$ (for some $m,n>0$) constructed from its corresponding synchronous closed loop system $\encap{\block}{P\merge_\gamma S}$ where,
\begin{itemize}
\item $\gamma':\left(\act\cup\hat{\act}\right)\times \left(\act\cup\hat{\act}\right)\rightarrow \left(\act\cup\hat{\act}\right)$ is the modified communication function (or the abstraction scheme for method M1) defined in following way,\newline
$\gamma'(!a,?\hat{a})=
\begin{cases}
\com\hat{a} \mbox{  if $!a\in\alpha^!(P)$}\\
\com a \mbox{  if $!a\in\alpha^!(S)$}
\end{cases}$
$\gamma'(!\hat{a},?a)=\begin{cases}
\com \hat{a} \mbox{  if $?a\in\alpha^?(P)$}\\
\com a \mbox{  if $?a\in\alpha^?(S)$}
\end{cases}.$

Intuitively, the communication function $\gamma'$ with the operators $\tau_{\hat{I}},\partial_{H\cup\hat{H}}$ ensures the interactions between the plant and the bag are invisible while the interactions between the supervisor and the bag are visible.
\end{itemize}

The rationale behind the choice of M1 is based on the observation that a transition system generated by a supervisor $S$ is isomorphic to the corresponding synchronous closed loop system $\encap{\block}{P \merge_\gamma S}$, modulo the difference in the type of action labels \citep{prDCL}. This is because in the synthesis of supervisors no transitions are introduced that a plant cannot execute. Moreover, the action labels in $S$ will be decorated as either an input action (?) or an output action (!) while in $\encap{\block}{P \merge_\gamma S}$ the same label will be decorated as a communicated action ($\com$). Formally, this fact is equivalent to
\[\rname{f}{S}\bisim \encap{\block}{P \merge_\gamma S}\]
where, $\rho$ is the renaming operator 
from TCP \cite{acpbook} and $f:\act\rightarrow\act$ is a function that renames an input/output action to a communicated action, i.e., $\forall ?a,!a\in\act.[f(!a)=f(?a)=\com a]$.
As a consequence, when one introduces bags and abstracts the interaction between plant and bags, the supervisor model remains unaffected. While in other abstraction schemes this is not the case. Thus, it is easier to study abstraction scheme M1 than other schemes. 

\section{Desynchronisable closed loop system}\label{sec:DCL}

In the previous section, we have shown how to construct an asynchronous closed loop system from a given synchronous closed loop system. In general, the newly constructed asynchronous closed loop system will not be branching bisimilar to the given synchronous closed loop system. To this end, we introduce a special class of the synchronous closed loop system called desynchronisable closed loop system that are always branching bisimilar to their corresponding asynchronous closed loop systems. We then present sufficient conditions for desynchronisability.

\newcommand{\bagmn}{\ensuremath{\bag^{m,n}}}

\begin{definition}\label{def-desyncl}
Let $\encap{\block}{P\merge_\gamma S}$ be a synchronous closed loop system and let $m,n$ be any two nonzero natural numbers. Then, $\encap{\block}{P\merge_\gamma S}$ is said to be \textit{desynchronisable} with input and output buffers of size $n$ and $m$ (or in short desynchronisable closed loop system), respectively, if $$\encap{\block}{P\merge_\gamma S}\bbisim \nabla{P \merge_{\gamma'} \bagmn[\ebag,\ebag] \merge_{\gamma'} S}\enspace.$$
\qed
\end{definition}

We now present three sufficient conditions for desynchronisability with buffers of arbitrary size. The objective of these conditions is the following. The conditions given in Definition~\ref{def-wellposed} and Definition~\ref{def-reordering} prevent an asynchronous closed loop system from getting deadlocked. The condition in Definition~\ref{def-diamond} ensures that the silent steps introduced by the abstraction scheme are inert.

\begin{definition}\label{def-wellposed}
Let $\encap{\block}{P\merge_\gamma S}$ be a synchronous closed loop system. Then, $\encap{\block}{P\merge_\gamma S}$ is called \textit{well posed} if there exists a binary relation $W\subseteq\prc\times\prc$ such that $(P,S)\in W$ and the following conditions are satisfied:
\begin{itemize}
\item $\forall\, !a,p,p',s.[(p,s)\in \well\wedge p\step{!a}p'\Rightarrow \exists s'.[s\step{?a}s'\wedge (p',s')\in\well]]\enspace,$ and
\item $\forall\, !a,p,s,s'.[(p,s)\in\well\wedge s\step{!a}s'\Rightarrow \exists p'.[p\step{?a}p'\wedge (p',s')\in\well]]\enspace.$
\end{itemize}
\qed
\end{definition}

\begin{figure}\centering
\includegraphics[width=0.4\textwidth,bb=14 14 347 153]{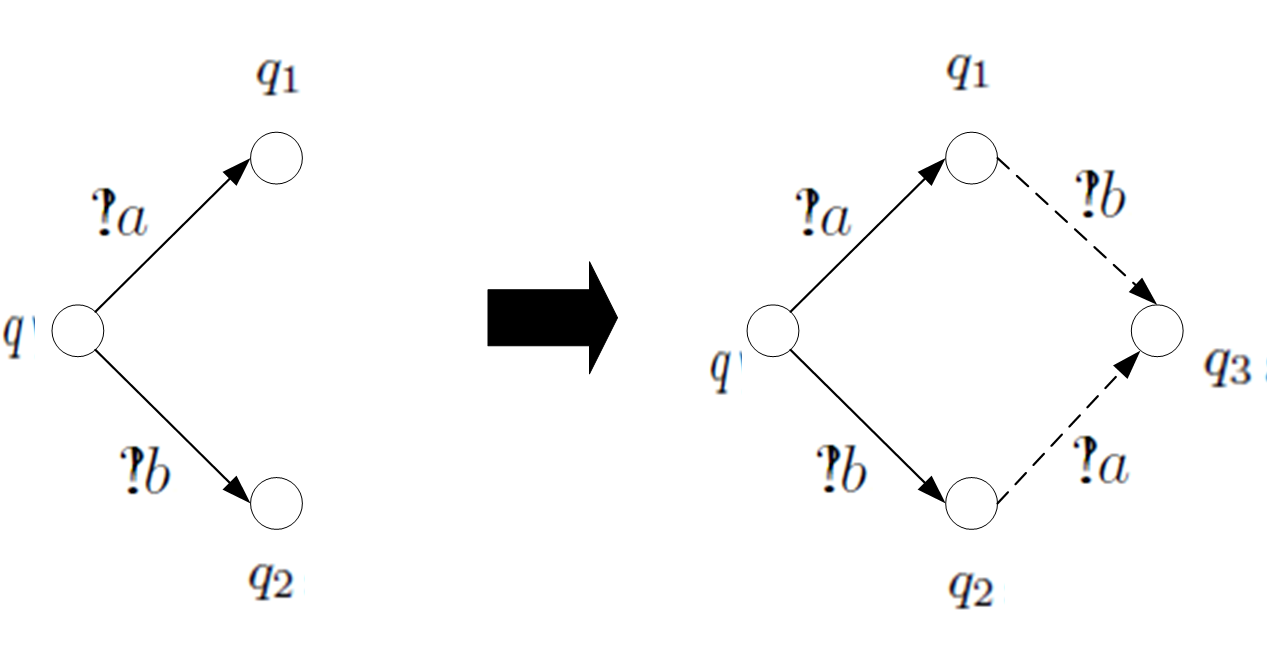}
\caption{Diamond property.}\label{diamond-fig}
\end{figure}

We now partition \label{partition} the set $\hide$ into two disjoint non-empty subsets $\hidepi,\hidepo$ with respect to a plant process $P$ as:
\begin{itemize}
\item $\hidepi\triangleq\{\com a\mid \com a\in \hide\wedge ?a\in\alpha^?(P)\}$.
\item $\hidepo\triangleq\{\com a\mid \com a\in \hide\wedge !a\in\alpha^!(P)\}$.
\end{itemize}

\begin{definition}\label{def-reordering}
Let $\mu\in\hidepi^{*}$ and $\nu\in\hidepo^{*}$ be sequences in $\hidepi$ and $\hidepo$, respectively. Let $p\in\prc$ be an arbitrary process and let $q\in Q$ be its initial state. We define the set of reachable states of $p$ in the following way,
    \begin{eqnarray*}
    &&\reach(q)=\{q'\mid\exists w\in\act^{*}.[q\steps{w}q']\}\enspace.
    \end{eqnarray*}
    By the semantics of TCP we know that if the initial state of a process is of the structure $\encap{\block}{\_\merge_\gamma \_}$ then, all the reachable states will also be of the same structure. A synchronous closed loop system $\encap{\block}{P\merge_\gamma S}$ is said to satisfy the \textit{reordering} property iff both the following conditions are satisfied,
    \begin{eqnarray*}
    &\bullet&\forall p',p_2,s',\encap{\block}{p_1\merge_\gamma s_1}\in\reach(\encap{\block}{P\merge_\gamma S}),\com a\in\hidepi.\\
&&\big[\encap{\block}{p_1\merge_\gamma s_1}\steps{\mu.\com a}\encap{\block}{p'\merge_\gamma s'}\wedge p_1\step{?a}p_2 \Rightarrow
\exists s_2.[\encap{\block}{p_1\merge_\gamma s_1}\step{\com a}\encap{\block}{p_2\merge_\gamma s_2}]\big]\\
    &\bullet&\forall p',s',s_2,\encap{\block}{p_1\merge_\gamma s_1}\in\reach(\encap{\block}{P\merge_\gamma S}),\com a\in\hidepo.\\
&&\big[\encap{\block}{p_1\merge_\gamma s_1}\steps{\nu.\com a}\encap{\block}{p'\merge_\gamma s'}\wedge s_1\step{?a}s_2\Rightarrow
\exists p_2.[\encap{\block}{p_1\merge_\gamma s_1}\step{\com a}\encap{\block}{p_2\merge_\gamma s_2}]\big].
    \end{eqnarray*}
\qed
\end{definition}

\begin{definition}\label{def-diamond}
Let $q\in Q$ be an arbitrary state. Then, $q$ is said to satisfy the \textit{diamond} property iff the following condition (see Figure~\ref{diamond-fig}) holds
\begin{itemize}
\item $\forall \com a,\com b\in\hide,q_1,q_2.\Big[q \step{\com a} q_1 \wedge q \step{\com b} q_2 \wedge \com a\neq\com b\Rightarrow \exists q_3.[q_1\step{\com b} q_3 \wedge q_2 \step{\com a} q_3]\Big]\enspace.$
\end{itemize}
A process $p$ is said to satisfy the diamond property iff for all reachable states $q'$ from $q$ satisfy the diamond property, where $q$ is the \emph{initial state} of the process $p$. \qed
\end{definition}




For a reader familiar with the concepts of \emph{true concurrency} \cite{Winskel95}, the conditions given in Definition~\ref{def:determinism}, \ref{def-reordering} and \ref{def-diamond} are similar to the axioms of asynchronous transition system. The formulation of these axioms is based on the definition of an independence relation, which is an irreflexive, symmetric relation on the set of actions $\act$. However, the techniques for desynchronisability for such models are not investigated here, although it will be worthwhile to examine this research direction in the future. Note that in our approach we do not need an additional notion of the independence relation.

Next, we present the following main results of this paper.
\begin{itemize}
\item If an arbitrary synchronous closed loop system satisfy the conditions in Definitions~\ref{def-wellposed}, \ref{def-reordering} and \ref{def-diamond} then, it is a desynchronisable  with buffers of arbitrary size.
\item The transition system generated by an asynchronous closed loop system constructed from a synchronous closed loop system satisfying the conditions in Definitions~\ref{def-wellposed}, \ref{def-reordering} and \ref{def-diamond} is always $\tau$-inert with respect to $\bbisim$.
\end{itemize}
To prove the above statements, we first fix some notations and then prove some lemmas, which are necessary for the proof of main theorem.

We denote the contents of an arbitrary bag by the symbols $\xi,\xi'$, i.e., $\xi,\xi'$ are of the form $(I_0,\kappa_0)$ and ($I_1,\kappa_1$) respectively, where $I_0,I_1\subseteq\hide$. The contents of the bag attached to input actions of $P$ is denoted by $\mu$, i.e., $\mu$ is of the form $(I_\mu,\kappa_\mu)$ where $I_\mu\subseteq\hidepi$. Similarly the contents of the bag attached to output actions of $P$ is denoted by $\nu$, i.e., $\nu$ is of the form $(I_\nu,\kappa_\nu)$ where $I_\nu\subseteq\hidepo$. For an arbitrary multiset $\xi$, we define a sequence (denoted as $\vec{\xi}$) over $\xi$ as,
$$\vec{\xi}\triangleq <x_1.x_2.\dots>$$ such that $\#(x_i,\vec{\xi})=\kappa(x_i)$, where $\#$ is a function that returns the maximum number of occurrences of $x_i$ in $\vec{\xi}$ for some $i>0$. For example consider a multiset $\xi=\{\com a,\com a,\com b,\com b\}$. Then a possible sequence $\vec{\xi}$ over the given $\xi$ can be of the form $<\com a.\com b.\com a.\com b>$. Let $f_i:{\hide}^{*}\rightarrow\block^{*}$ be the function defined as $f_i(\com a.\vec{\xi})=?a.f_i(\vec{\xi})$. Similarly, let $f_o:{\hide}^{*}\rightarrow\block^{*}$ be the function defined as $f_o(\com a.\vec{\xi})=!a.f_o(\vec{\xi})$.

\begin{proposition}\label{prop:mu'}
Given a trace $\encap{\block}{P\merge_\gamma S}\steps{\vec{\mu}}\encap{\block}{p_1\merge_\gamma s_1}$, we find using the above function $f_i$ and semantics of $\merge_\gamma$ that $P\steps{f_i(\vec{\mu})}p_1\wedge S\steps{f_o(\vec{\mu})}s_1$.
\end{proposition}
\begin{proposition}\label{prop:mu}
Similarly, given a trace $\encap{\block}{P\merge_\gamma S}\steps{\vec{\nu}}\encap{\block}{p_1\merge_\gamma s_1}$, we conclude that $P\steps{f_o(\vec{\nu})}p_1\wedge S\steps{f_i(\vec{\nu})}s_1$.
\end{proposition}

The following lemma is a generalisation of Definition~\ref{def-diamond}. It states that if two different states $q_1,q_2$ are reachable from a state $q_0$, then there exists a state $q_3$ reachable from $q_1$ and $q_2$ such that, the trace between $q_0,q_1$ and the trace between $q_0,q_2$ commute.

\begin{lemma}[\textbf{Generalised diamond property}]\label{gdp}
Let $\encap{\block}{P\merge_\gamma S}$ be an arbitrary synchronous closed loop system satisfying the conditions in Definitions~\ref{def-wellposed}, \ref{def-reordering} and \ref{def-diamond}. If $\encap{\block}{P\merge_\gamma S}\steps{\vec{\xi}}\encap{\block}{p_1\merge_\gamma s_1}\; \wedge\; \encap{\block}{P\merge_\gamma S}\steps{\vec{\xi'}}\encap{\block}{p_2\merge_\gamma s_2}$ then,
    \[\exists p_3,s_3.[\encap{\block}{p_1\merge_\gamma s_1}\steps{\vec{\xi'}}\encap{\block}{p_3\merge_\gamma s_3} \wedge \encap{\block}{p_2\merge_\gamma s_2}\steps{\vec{\xi}}\encap{\block}{p_3\merge_\gamma s_3}].\]
\end{lemma}

The following Lemmas~\ref{dppi}, \ref{dppo} are the results (See \citep{desync} for the proofs) obtained by direct instantiation of reordering property (Definition~\ref{def-reordering}) and generalised diamond property (Lemma~\ref{gdp}).

\begin{lemma}\label{dppi}
Let $\encap{\block}{P\merge_\gamma S}$ be a synchronous closed loop system satisfying the conditions in Definitions~\ref{def-wellposed}, \ref{def-reordering} and \ref{def-diamond}. Suppose $\com a\in\hidepi \wedge\, \encap{\block}{P\merge_\gamma S}\steps{\vec{\mu}.\com a}\encap{\block}{p_2\merge_\gamma s_2}\wedge P\step{?a}p_1$ then, \begin{eqnarray*}
    &&\exists s_1.\left[\encap{\block}{P\merge_\gamma S}\step{\com a}\encap{\block}{p_1\merge_\gamma s_1}\steps{\vec{\mu}}\encap{\block}{p_2\merge_\gamma s_2}\right].\end{eqnarray*}
\end{lemma}

\begin{lemma}\label{dppo}
Let $\encap{\block}{P\merge_\gamma S}$ be a synchronous closed loop system satisfying the conditions in Definitions~\ref{def-wellposed}, \ref{def-reordering} and \ref{def-diamond}. Suppose $\com a\in\hidepo\wedge \,\encap{\block}{P\merge_\gamma S}\steps{\vecn.\com a}\encap{\block}{p_3\merge_\gamma s_3}\wedge S\step{?a}s_1$ then, \begin{eqnarray*}
    &&\exists p_1.\left[\encap{\block}{P\merge_\gamma S}\step{\com a}\encap{\block}{p_1\merge_\gamma s_1}\steps{\vecn}\encap{\block}{p_3\merge_\gamma s_3}\right].\end{eqnarray*}
\end{lemma}

We now pose the main theorem of this paper which proves the following statement. If the given synchronous closed system satisfies the conditions in Definition~\ref{def-wellposed}, \ref{def-reordering} and \ref{def-diamond}, then it is desynchronisable independent of the size of the buffers introduced between the given plant and its supervisor. 

\begin{theorem}\label{thm:char}
Let $\encap{\block}{P\merge_\gamma S}$ be an arbitrary synchronous closed loop system satisfying the conditions in Definitions~\ref{def-wellposed}, \ref{def-reordering} and \ref{def-diamond}. Then for any $m,n>0$ we have, $$\encap{\block}{P\merge_\gamma S}\bbisim\nabla{P\merge_{\gamma'}\bagmn[\varepsilon,\varepsilon\,]
\merge_{\gamma'}S}.$$
\begin{proof}
Let $p,p',s$ be free process variables. Let $\mu,\nu$ be two free variables representing the contents of an input and an output bag of $P$, respectively. Then, define a relation $\Phi$ as follows.
\begin{eqnarray*}
\begin{split}
\Phi\triangleq\{(\encap{\block}{p\merge_\gamma s},\nabla{p'\merge_{\gamma'}\bag^{m,n}[\mu,\nu]\merge_{\gamma'}s}) \mid\; \\
\Big(p'=p\wedge\mu=\varepsilon\wedge\nu=\varepsilon\Big)&\bigvee \;\;\mbox{ [C1]}\\
\Big(\mu=\varepsilon\wedge \exists s'.\big[\encap{\block}{p\merge_\gamma s}\steps{\vecn}\encap{\block}{p'\merge_\gamma s'}\big]\Big)&\bigvee \;\;\mbox{ [C2]}\\
\Big(\nu=\varepsilon\wedge \exists s'.\big[\encap{\block}{p'\merge_\gamma s'}\steps{\vecm}\encap{\block}{p\merge_\gamma s}\big]\Big)&\bigvee \;\;\mbox{ [C3]}\\
\Big(\exists p'',s',s''.\big[\encap{\block}{p'\merge_\gamma s'}\lla{\vecn}{}\encap{\block}{p''\merge_\gamma s''}\steps{\vecm}\encap{\block}{p\merge_\gamma s}\big]\Big)&\bigvee \;\;\mbox{ [C4]}\\
\Big(\exists p'',s',s''.\big[\encap{\block}{p\merge_\gamma s}\steps{\vecn}\encap{\block}{p''\merge_\gamma s''}\lla{\vecm}{}\encap{\block}{p'\merge_\gamma s'}\big]\Big)&\enspace\}.\;\;\mbox{ [C5]}\end{split}
\end{eqnarray*}
Note that the above conditions C1, C2, C3, C4 and C5 are independent of $n,m$. The proof of the theorem is based on showing that $\Phi$ is a witnessing branching bisimulation relation. The intuition behind the definition of $\Phi$ is the following. A state $\encap{\block}{p \merge_\gamma s}$ in a synchronous closed loop system is related to those states in an asynchronous closed loop system that contain the same supervisor state $s$. The $\Phi$ relation between two states is indicated by dotted lines in Figure~\ref{fig:phi}.
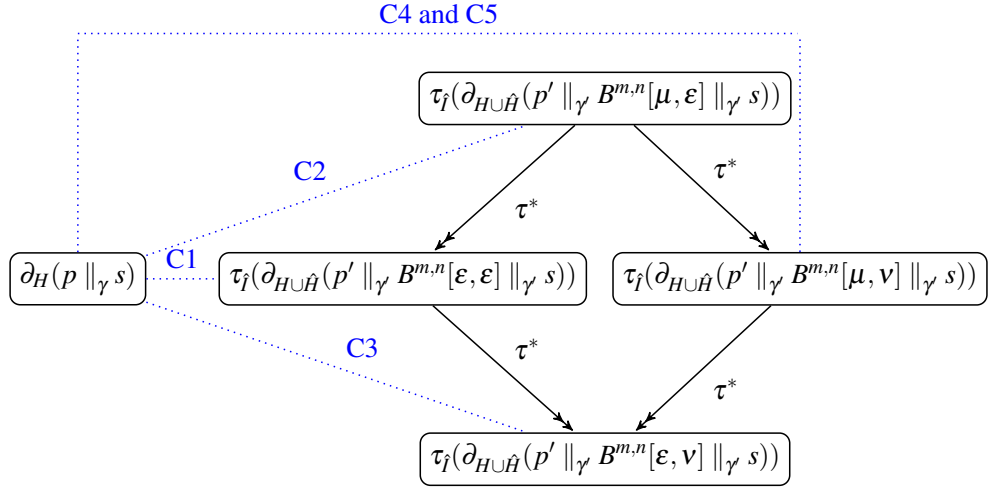
\begin{figure}
        \centering
        \scalebox{0.96}{
        \begin{tikzpicture}[->,>=stealth',shorten >=1pt,auto, semithick,]

  \tikzset{
    state/.style={
      circle,
      draw=black,
    }
  }

  \tikzset{
    super/.style={
      rectangle,
      rounded corners,
      draw=black,
    }
  }

\node[super] (qc) {$\encap{\block}{p\merge_\gamma s}$};
\node[super] (qac) at ($(qc.center) + (4.5,0)$) {$\nabla{p'\merge_{\gamma'}\bag^{m,n}[\ebag,\ebag]\merge_{\gamma'}s}$};
\node[super] (qacf) at ($(qac.center) + (5.5,0)$)
{$\nabla{p'\merge_{\gamma'}\bag^{m,n}[\mu,\nu]\merge_{\gamma'}s}$};
\node[super] (qact) at ($(qac.center) + (2.8,2.5)$)
{$\nabla{p'\merge_{\gamma'}\bag^{m,n}[\mu,\ebag]\merge_{\gamma'}s}$};
\node[super] (qacd) at ($(qac.center) + (2.8,-2.5)$)
{$\nabla{p'\merge_{\gamma'}\bag^{m,n}[\ebag,\nu]\merge_{\gamma'}s}$};
\node (eqct) at ($(qc.center) + (0,3.4)$) {};
\node (eqacft) at ($(qacf.center)+(0,3.4)$) {};

\path[-,dotted,blue] (qc) edge node {C1} (qac);
\path[-,dotted,blue] (qc) edge node {C2} (qact);
\path[-,dotted,blue] (qc) edge node {C3} (qacd);
\draw[-,dotted,blue] (qc) -- ($(qc.center) + (0,3.4)$);
\draw[-,dotted,blue] (qacf) -- ($(qacf.center) + (0,3.4)$);
\path[-,dotted,blue] ($(qc.center) + (0,3.4)$) edge node {C4 and C5} ($(qacf.center) + (0,3.4)$);

\path[->>] (qact) edge node {$\tau^{*}$} (qac);
\path[->>] (qact) edge node {$\tau^{*}$} (qacf);
\path[->>] (qacf) edge node {$\tau^{*}$} (qacd);
\path[->>] (qac) edge node {$\tau^{*}$} (qacd);

\end{tikzpicture} 
        }
\caption{Illustration of relation $\Phi$.}\label{fig:phi}
\end{figure}
The complete proof requires a lot of case distinction and can be found in \citep{desync}. Here we discuss the different cases that are present in the proof and give the list of lemmas that are applied in each case. Let $q_c,q_a$ be the initial states of the processes $\encap{\block}{p\merge_\gamma s}$ and $\nabla{p'\merge_{\gamma'}\bag^{m,n}[\mu,\nu]\merge_{\gamma'}s}$, respectively. From the definition of branching bisimilarity we need to show the following four transfer conditions:
\begin{enumerate}
\item $\forall \com a,q_c,q_c',q_a.[q_c\step{\com a}q_c'\wedge (q_c,q_a)\in\Phi\Rightarrow \exists q_a',q_a''.[q_a\steps{\tau^{*}}q_a'\step{\com a}q_a'' \wedge (q_c,q_a'),(q_c',q_a'')\in\Phi]].$
\item\label{cond2} $\forall q_c,q_c',q_a.[q_c\step{\tau}q_c'\wedge (q_c,q_a)\in\Phi\Rightarrow (q_c',q_a)\in\Phi\vee\exists q_a',q_a''.[q_a\steps{\tau^{*}}q_a'\step{\tau}q_a'' \wedge (q_c,q_a'),(q_c',q_a'')\in\Phi]].$
\item $\forall q_c,q_a,q_a'.[q_a\step{\tau}q_a' \wedge (q_c,q_a)\in\Phi\Rightarrow (q_a',q_c)\in\Phi\vee\exists q_c',q_c''.[q_c\step{\tau^{*}}q_c'\step{\tau}q_c''\wedge (q_c',q_a),(q_c'',q_a')\in\Phi]]$.
\item $\forall \com a,q_c,q_a,q_a'.[q_a\step{\com a}q_a'\wedge (q_c,q_a)\in\Phi\Rightarrow \exists q_c',q_c''.[q_c\step{\tau^{*}}q_c'\step{\com a}q_c''\wedge (q_c',q_a),(q_c'',q_a')\in\Phi]]$.
\end{enumerate}

Since the synchronous closed loop system does not contain $\tau$ actions in its alphabet, there are following three effects on the above transfer conditions.
\begin{itemize}
\item The above condition 2 will be vacuously satisfied.
\item The condition 3 will be reduced to the simpler form,
\begin{eqnarray*}
&&\forall q_c,q_a,q_a'.\big[q_a\step{\tau}q_a' \wedge (q_c,q_a)\in\Phi\Rightarrow (q_c,q_a')\in\Phi\big].
\end{eqnarray*}
\item And similarly condition 4 will be reduced to:
\begin{eqnarray*}
&&\forall \com a,q_c,q_a,q_a'.[q_a\step{\com a}q_a'\wedge (q_c,q_a)\in\Phi\Rightarrow \exists q_c'.[q_c\step{\com a}q_c'\wedge (q_c',q_a')\in\Phi]].
\end{eqnarray*}
\end{itemize}
But to show that these conditions hold, we need to know whether an action label $\com a$ occurring in each condition is either an input or output action with respect to $P$, i.e. $\com a\in\hidepi$ or $\com a\in\hidepo$. 
Thus, we get six transfer conditions in total that are shown in Table~\ref{table1}. Furthermore, for each case we apply case distinction based on the structure of $\mu$ and $\nu$. In each subcase we use C1, C2, C3, C4, and C5 (the conditions from the definition of $\Phi$) to determine the relation between free process variables $p,p'$ and then prove the conclusion as shown in Table~\ref{table1}. The notation $\tau=\abstr{\com a}{\com a}$ is used to denote that the $\tau$ action is a result of abstraction of the communicated action $\com a$.
\begin{table}[ht]
\caption{Proof structure of Theorem~\ref{thm:char}.}
\centering
\begin{tabular}{|c|l|l|}
  \hline
  Case No. & Hypothesis & Conclusion\\
  \hline
  \inct & $q_c\step{\com a}q_c'\wedge (q_c,q_a)\in\Phi\wedge \com a\in\hidepi$. & $\exists q_a',q_a''.[q_a\steps{\tau^{*}}q_a'\step{\com a}q_a'' \wedge (q_c,q_a'),(q_c',q_a'')\in\Phi.]$\\
  \inct & $q_c\step{\com a}q_c'\wedge (q_c,q_a)\in\Phi\wedge \com a\in\hidepo$.& $\exists q_a',q_a''.[q_a\steps{\tau^{*}}q_a'\step{\com a}q_a'' \wedge (q_c,q_a'),(q_c',q_a'')\in\Phi.]$\\
  \inct & $q_a\step{\tau}q_a'\wedge (q_c,q_a)\in\Phi \wedge \com a\in\hidepi$& $(q_c,q_a')\in\Phi.$ \\
  & $\quad \wedge\ \tau=\abstr{\com a}{\com a}$. & \\
  \inct & $q_a\step{\tau}q_a'\wedge (q_c,q_a)\in\Phi \wedge \com a\in\hidepo$& $(q_c,q_a')\in\Phi.$ \\
  & $\quad \wedge\ \tau=\abstr{\com a}{\com a}$. &\\
  \inct & $q_a\step{\com a}q_a'\wedge (q_c,q_a)\in\Phi\wedge \com a\in\hidepi$.& $\exists q_c'.[q_c\step{\com a}q_c'\wedge (q_c',q_a')\in\Phi.]$ \\
  \inct & $q_a\step{\com a}q_a'\wedge (q_c,q_a)\in\Phi\wedge \com a\in\hidepo$.&$\exists q_c'.[q_c\step{\com a}q_c'\wedge (q_c',q_a')\in\Phi.]$ \\
  \hline
\end{tabular}
\label{table1}
\end{table}

In Table~\ref{lemma-applicable} we present the list of lemmas required to prove each case.

\begin{table}
  \centering
  \begin{tabular}{|l|l|}
    \hline
    Case No. & List of lemmas\\
    \hline
    T1 & Lemma~\ref{gdp} \\
    T2 & Lemma~\ref{gdp} and Proposition~\ref{prop:mu'} \\
    T3 & Lemma~\ref{gdp} and Lemma~\ref{dppi}\\
    T4 & Lemma~\ref{gdp} \\
    T5 & Lemma~\ref{gdp} and Lemma~\ref{dppo} \\
    \hline
  \end{tabular}
  \caption{List of lemmas applied in each case.}\label{lemma-applicable}
\end{table}

\end{proof}
\end{theorem}

In hindsight, what we have actually proven is that all $\tau$ actions generated by the abstraction scheme are $\tau$-inert with respect to $\bbisim.$ The following corollary states this fact.

\begin{corollary}\label{thm:tauinert}
Let $q_c$ be a process of the form $\encap{\block}{P\merge_\gamma S}$. And let $q_a$ be an asynchronous closed loop system of the form $\nabla{P'\merge_{\gamma'}\bagmn[\mu,\nu]\merge_{\gamma'}S}$ such that $(q_c,q_a)\in\Phi$. Then,
\[\forall q_c,q_a,q_a'.\big[(q_c,q_a)\in\Phi\wedge q_a\step{\tau}q_a'\Rightarrow q_a\bbisim q_a'\big].\]
\end{corollary}

\section{Conclusions and future work}\label{sec:discussion}

In this paper, we presented sufficient conditions for desynchronisability in a process algebraic setting and showed that an asynchronous implementation using bags (of arbitrary size) is a refinement of the synchronous closed loop system satisfying these conditions.
The prominent features of our work can be summarised in the following main points:
\begin{itemize}
\item We solve a refinement problem instead of a supervisory control problem, and do not compute a new supervisor in the presence of buffers, as done in \citep{balemiphdt,async-imp}. Our approach is \textit{intended} to be computationally cheaper than the one developed in \citep{balemiphdt,async-imp}, however this conjecture needs to be verified by analysing the complexities associated with the conditions presented here. In particular, we conjecture that supervisory control theory always results in synchronous closed loop systems, which are well-posed (Definition~\ref{def-wellposed}), but the other conditions, (Definition~\ref{def-reordering} and Definition~\ref{def-diamond}), are not likely to be attained so easily.
\item We present our conditions for desynchronisability over the components of a synchronous closed loop system conjointly, in contrast with \citep{Fischer96}, where the check for the foam rubber wrapper principle on the two components was applied separately. Note the sender domination property from \citep{Fischer96} is equivalent to the well posed condition (Definition~\ref{def-wellposed}). However, the two approaches are incomparable because in \citep{Fischer96} the construction method M3 was studied while in this paper the construction method M1 is studied.
\item We use branching bisimulation equivalence instead of failure equivalence that was adopted in \citep{Fischer96}. As a consequence our techniques are applicable to all the weak equivalences in `van Glabbeek spectrum' \citep{Glabeek90} (including failure equivalence). The branching bisimulation is the preferred equivalence in TCP process algebra under the presence of $\tau$ action \citep{acpbook}. 
    Furthermore, the conditions (well posedness and diamond property) given here are similar to the ones mentioned in \citep{Fischer96}, where desynchronisability was characterised modulo failure equivalence. Thus, we conjecture that achieving desynchronisability for weaker equivalences will not lead to weaker sufficient conditions.
\end{itemize}

A question that was not treated in this paper, is whether the conditions we posed are in fact reasonable for industrial applications. This may become clear in the near future, when we
study the case studies involved with supervisory control theory in the context of MULTIFORM project \citep{multiform} with the language CIF \citep{cif}. The authors of CIF are currently developing techniques that will incorporate supervisory control theory and model based engineering into a single framework, thus making it suitable for the design of industrial applications. In particular, the elevator case study and the toy example, which were desynchronisable in \citep{prDCL} using the construction method M1, satisfy our conditions.

Another question is whether the conditions given are actually necessary for desynchronisability modulo branching bisimilarity. Formally speaking, they are not, and counter-examples have been found although we do not give them here. We anticipate that the diamond property (Definition~\ref{def-diamond}) can be further weakened. In particular, if the actions $\com a,\com b\in\hidepi$ are enabled at a state $q$ then it may not be necessary for the traces $\com a.\com b$ and $\com b.\com a$ to commute.

Lastly, the research performed in this paper can of course be repeated for different architectures. One might study whether wires or queues can be used instead of bags, or study different abstraction schemes, or try to study the conditions for desynchronisability by focusing on other notions of weak equivalences.

\section{Acknowledgements}\label{sec:ack}
The authors would like to thank the reviewers for their valuable critical remarks on the earlier draft of this paper and pointing out that the sender domination property and the well posed property are equivalent. The authors would also like to thank Jos Baeten, Koos Rooda, Bert van Beek,
Rong Su, Jasen Markowski, Damian Nadales, and Mihaly Petreczky for various discussions and clarification about this work.

This work has been performed as part of the ``Integrated Multi-formalism Tool Support for the Design of Networked Embedded Control Systems'' (MULTIFORM) project, supported by the Seventh Research Framework Programme of the European Commission (Grant agreement number: INFSO-ICT-224249).

\bibliographystyle{plainnat}
\bibliography{paper}

\begin{thebibliography}{18}
\providecommand{\natexlab}[1]{#1}
\providecommand{\url}[1]{\texttt{#1}}
\expandafter\ifx\csname urlstyle\endcsname\relax
  \providecommand{\doi}[1]{doi: #1}\else
  \providecommand{\doi}{doi: \begingroup \urlstyle{rm}\Url}\fi

\bibitem[mul()]{multiform}
Integrated multi-formalism tool support for the design of networked embedded
  control systems : Multiform.
\newblock \url{http://cms.multiform.bci.tu-dortmund.de/}.

\bibitem[Agut et~al.(2009)Agut, van Beek, Schiffelers, Hendriks, and
  Rooda]{cif}
D.~E.~Nadales Agut, D.~A. van Beek, R.~R.~H. Schiffelers, D.~Hendriks, and
  J.~E. Rooda.
\newblock Abstract syntax and formal semantics of the {CIF}.
\newblock Technical report, Eindhoven {U}niversity of {T}echnology, October
  2009.

\bibitem[Baeten et~al.(2009)Baeten, Basten, and Reniers]{acpbook}
J.~C.~M. Baeten, T.~Basten, and M.~Reniers.
\newblock \emph{Process Algebra: Equational Theories of Communicating
  Processes}.
\newblock Cambridge University Press, 2009.

\bibitem[Balemi(1992)]{balemiphdt}
S.~Balemi.
\newblock \emph{Control of Discrete Event Systems: Theory And Application}.
\newblock PhD thesis, Swiss Federal Institute of Technology, Automatic Control
  Laboratory, ETH Zurich, May 1992.

\bibitem[Beohar and Cuijpers(2010)]{desync}
H.~Beohar and P.~J.~L. Cuijpers.
\newblock A theory of desynchronisable closed loop systems.
\newblock To appear as {T}echnical report, Eindhoven university of technology,
  2010.
\newblock \url{http://www.win.tue.nl/~pcuijper/pages/publications.html}.

\bibitem[Beohar et~al.(2009)Beohar, Cuijpers, and Baeten]{prDCL}
H.~Beohar, P.~J.~L. Cuijpers, and J.~C.~M. Baeten.
\newblock Design of asynchronous supervisors.
\newblock abs/0910.0868, 2009.
\newblock \url{http://arxiv.org/abs/0910.0868}.

\bibitem[Fabian and Hellgren(1998)]{fabian}
M.~Fabian and A.~Hellgren.
\newblock {PLC}-based implementation of supervisory control for discrete event
  systems.
\newblock \emph{Proceedings of the 37th IEEE Conference on Decision and
  Control}, 3:\penalty0 3305--3310, 1998.

\bibitem[Fischer and Janssen(1996)]{Fischer96}
C.~Fischer and W.~Janssen.
\newblock Synchronous development of asynchronous systems.
\newblock In Ugo Montanari and Vladimiro Sassone, editors, \emph{Proceedings of
  CONCUR'96}, volume 1119 of \emph{Lecture Notes in Computer Science}, pages
  735--750. Springer-Verlag, 1996.

\bibitem[Groote and Sellink(1996)]{confluence}
J.~F. Groote and M.~P.~A. Sellink.
\newblock {C}onfluence for process verification.
\newblock \emph{Theor. Comput. Sci.}, 170\penalty0 (1-2):\penalty0 47--81,
  1996.

\bibitem[{He Jifeng} et~al.(1990){He Jifeng}, Josephs, and Hoare]{HHJ90}
{He Jifeng}, M.~B. Josephs, and C.~A.~R. Hoare.
\newblock A theory of synchrony and asynchrony.
\newblock In M.~Broy and C.~B. Jones, editors, \emph{Programming Concepts and
  Methods}, pages 459--479, 1990.

\bibitem[Kapoor and Josephs(2004)]{diccs}
H.~K. Kapoor and M.~B. Josephs.
\newblock Modelling and verification of delay-insensitive circuits using {CCS}
  and the concurrency workbench.
\newblock volume~89, pages 293--296, 2004.

\bibitem[Mousavi et~al.(2003)Mousavi, Le~Guernic, Talpin, Shukla, and
  Basten]{MousaviDate04}
M.~Mousavi, P.~Le~Guernic, J.-P. Talpin, S.~K. Shukla, and T.~Basten.
\newblock Modeling and validating globally asynchronous design in synchronous
  frameworks.
\newblock In \emph{Proceedings of the Conference on Design Automation and Test
  in Europe}, pages 384--389. IEEE Computer Society Press, 2003.

\bibitem[Plotkin(1981)]{SOS}
G.~D. Plotkin.
\newblock {A Structural Approach to Operational Semantics}.
\newblock Technical Report DAIMI FN-19, University of Aarhus, 1981.

\bibitem[Ramadge and Wonham(1987)]{RW:1987}
P.~J. Ramadge and W.~M. Wonham.
\newblock Supervisory control of a class of discrete event processes.
\newblock \emph{SIAM Journal on Control and Optimization}, 25\penalty0
  (1):\penalty0 206--230, 1987.

\bibitem[Udding(1984)]{udding}
J.T. Udding.
\newblock \emph{Classification and Composition of Delay-Insensitive Circuits}.
\newblock PhD thesis, Eindhoven University of Technology, Eindhoven, 1984.

\bibitem[v.~Glabbeek(1990)]{Glabeek90}
R.~J. v.~Glabbeek.
\newblock \emph{Comparative concurrency semantics and refinement of actions}.
\newblock Ph.{D}. thesis, CWI, Amsterdam, 1990.

\bibitem[Winskel and Nielsen(1995)]{Winskel95}
G.~Winskel and M.~Nielsen.
\newblock Models for concurrency.
\newblock In \emph{Handbook of Logic in Computer Science}, pages 1--148. Oxford
  University Press, 1995.

\bibitem[Xu and Kumar(2008)]{async-imp}
S.~Xu and R.~Kumar.
\newblock Asynchronous implementation of synchronous discrete event control.
\newblock pages 181 --186, May 2008.
\newblock \doi{10.1109/WODES.2008.4605942}.

\end{thebibliography}
\appendix
\section{Operational semantics of TCP}\label{osrules}

In this section, we give the SOS rules for the operators used in this paper. Note that the rules for symmetric case (in the context of binary operators) are not given.
\[ 
\begin{array}{@{}c@{}}
\osrule{ }{x.p\step{x}p} \ \ \ \

\osrule{p\step{x}p'}{\begin{array}{c}
p\altc q\step{x} p' \\ q\altc p\step{x} p'
\end{array}} \ \ \ \

\osrule{p\step{x}p'}{\begin{array}{c}
p\merge_\gamma q \step{x} p'\merge_\gamma q \\ q\merge_\gamma p \step{x} q\merge_\gamma p'
\end{array}} 

\osrule{p\step{x}p',q\step{x'}q',\gamma(x,x')=x''}{\begin{array}{c}
p\merge_\gamma q \step{x''} p'\merge_\gamma q' \\
q\merge_\gamma p \step{x''} q' \merge_\gamma p'
\end{array}} \\
\\
\osrule{p\step{x}p',x\not\in H}{\encap{H}{p}\step{x}\encap{H}{p'}} \ \ \ \

\osrule{p\step{x}p',x\not\in I}{\abstr{I}{p}\step{x}\abstr{I}{p}} \ \ \ \

\osrule{p\step{x}p',x\in I}{\abstr{I}{p}\step{\tau}\abstr{I}{p}} \\
\\
\osrule{t_0\step{x}p, X_0=t_0}{X_0\step{x}p}\ \ \ \

\osrule{p \step{x} p',f:\act\rightarrow\act}{\rname{f}{p}\step{f(x)} \rname{f}{p'}}
\end{array}
\]

%
%
%
%
%
%
%
%
%
%

\begin{definition}
The alphabet of a process $p$, written as $\alpha(p)$, is the set of atomic actions that it can perform. It is defined for the following closed terms.
\begin{eqnarray*}
\alpha(\dl)&=&\emptyset\\
\alpha(x.p)&=&\{x\}\cup \alpha(p)\\
\alpha(\tau.p)&=&\alpha(p)\\
\alpha(p\altc q) &=& \alpha(p)\cup\alpha(q)
\end{eqnarray*}
Note that $\alpha$ is not defined explicitly for the operators $\merge_\gamma,\partial_H,\tau_I$ because these operators can be eliminated (See the corresponding elimination theorems in \citep{acpbook}).\qed
\end{definition}

\end{document}